\documentclass[aps,twocolumn,pre,showpacs,superscriptaddress,longbibliography]{revtex4-2}
\usepackage{bm}
\usepackage{amsmath}
\usepackage{amssymb}
\usepackage{color}
\usepackage{graphicx,dcolumn,multirow,amsmath,amssymb}
\usepackage[colorlinks,linkcolor=blue,citecolor=blue,anchorcolor=blue,urlcolor=blue]{hyperref}

\begin{document}
\newcommand{\blue}[1]{\textcolor{blue}{#1}}
\newcommand{\red}[1]{\textcolor{red}{#1}}
\newcommand{\green}[1]{\textcolor{green}{#1}}
\newcommand{\orange}[1]{\textcolor{orange}{#1}}

\def\today{November 13, 2025; revised February 7, 2026}
\title{Backbone three-point correlation function in the two-dimensional Potts model}

\author{Ming Li}
\email{lim@hfut.edu.cn}
\affiliation{School of Physics, Hefei University of Technology, Hefei 230009, China}

\author{Youjin Deng}
\email{yjdeng@ustc.edu.cn}
\affiliation{Hefei National Research Center for Physical Sciences at the Microscale, University of Science and Technology of China, Hefei 230026, China}
\affiliation{Department of Modern Physics, University of Science and Technology of China, Hefei 230026, China}
\affiliation{Hefei National Laboratory, University of Science and Technology of China, Hefei 230088, China}

\author{Jesper Lykke Jacobsen}
\email{jesper.jacobsen@ens.fr}
\affiliation{CNRS-Laboratoire de Physique de l'Ecole Normale Sup\'eriure, PSL Research University, Sorbonne Universit\'e, Universit\'e Paris Cit\'e,
24 rue Lhomond, 75005 Paris, France}
\affiliation{Sorbonne Universit\'e, Ecole Normale Sup\'eriure, CNRS, Laboratoire de Physique (LPENS)}
\affiliation{Institut de Physique Th\'eorique, CEA, CRNS, Universit\'e Paris--Saclay}

\author{Jes\'us Salas}
%%%\thanks{Retired}
\email{u3985545049@gmail.com}
\affiliation{Universidad Carlos III de Madrid, Legan\'es, Spain}

\date{\today}

\begin{abstract}
We study the three-point correlation function of the backbone in the two-dimensional $Q$-state Potts model using the Fortuin--Kasteleyn (FK) representation. The backbone is defined as the biconnected skeleton of an FK cluster after removing all dangling ends and bridges. To circumvent the severe critical slowing down in direct Potts simulations for large $Q$, we employ large-scale Monte Carlo simulations of the O$(n)$ loop model on the hexagonal lattice, which is regarded to correspond to the Potts model with $Q=n^2$. Using a highly efficient cluster algorithm, we compute the universal three-point amplitude ratios for the backbone ($R_\text{BB}$) and FK clusters ($R_\text{FK}$). Our computed $R_\text{FK}$ exhibits excellent agreement with exact conformal field theory predictions, validating the reliability of our numerical approach. In the critical regime, we find that $R_\text{BB}$ is systematically larger than $R_\text{FK}$. Conversely, along the tricritical branch, $R_\text{BB}$ and $R_\text{FK}$ coincide within numerical accuracy, strongly suggesting that $R_\text{BB}=R_\text{FK}$ holds throughout this regime. This finding mirrors the known equality of the backbone and FK cluster fractal dimensions at tricriticality, jointly indicating that both structures share the same geometric universality.
\end{abstract}

\maketitle

\section{Introduction}  \label{sec:intro}

The $Q$-state Potts model is one of the most extensively studied models in statistical physics, providing a unifying framework for understanding both continuous and discontinuous phase transitions~\cite{Potts_52,Wu_82,Wu_82a,Wu_84,Baxter_85}. In this model, spins interact via a ferromagnetic nearest-neighbor coupling, and the Hamiltonian (reduced by the Boltzmann constant $k_{\rm B}$ and temperature $T$) is given by
\begin{equation}
\mathcal{H}/k_{\rm B}T = -K \sum_{\langle ij \rangle} \delta_{\sigma_i,\sigma_j}\,,
\end{equation}
where the spin variable $\sigma_i$ takes one of $Q$ possible states, and $\delta_{a,b}$ is the Kronecker delta function.

Through the Fortuin--Kasteleyn (FK) representation~\cite{Kasteleyn_69,Fortuin_72}, the partition function of Potts model can be expressed in terms of random clusters as
\begin{equation}
\mathcal{Z}_{\text{FK}} = \sum_{\{\mathcal{E}\}} v^{b(\mathcal{E})} Q^{c(\mathcal{E})}\,, \qquad v = e^{K}-1\,,   \label{eq:zrc}
\end{equation}
where the sum runs over all possible subsets $\mathcal{E}$ of occupied bonds, $b(\mathcal{E})$ is the number of occupied bonds, and $c(\mathcal{E})$ is the number of connected components (FK clusters). This formulation provides a geometrical interpretation of the Potts model in terms of clusters (for $Q,v>0$), allowing its critical properties to be studied in close analogy with percolation theory~\cite{Stauffer_94,Grimmett_99,Bollobas_06}. Indeed, in the limit $Q\to1$, the partition function \eqref{eq:zrc} reduces to that of standard bond percolation. More generally, for arbitrary values of $Q$ and $v$, Eq.~\eqref{eq:zrc} defines the random-cluster model~\cite{Grimmett_06}. The FK representation has an important role in conformal field theory (CFT)~\cite{DiFrancesco_97} and in stochastic Loewner evolution~\cite{Kager_04,Cardy2005}, leading to major theoretical advances for the Potts model.

In the two-dimensional (2D) Potts model, the Coulomb-gas (CG) theory~\cite{Nienhuis_82,Nienhuis_84,Nienhuis_87} provides a unified theoretical framework. Within this theory, the critical exponents for different $Q$ can be consistently expressed as functions of the CG coupling $g$, which is related to $Q$ by
\begin{equation}
\sqrt{Q} = -2\cos(\pi g)\,.  \label{eq:Q_vs_g}
\end{equation}
The range $Q \in [0,4]$ corresponds to $g \in [1/2,1]$. 
In this framework, the thermal and magnetic renormalization exponents for the leading fields are given by
\begin{align}
y_{t1} & = 3-\frac{3}{2g}\,,   \label{eq-yt1}  \\
y_{h1} & = \frac{(2g+3)(2g+1)}{8g}\,,    \label{eq-yh1} 
\end{align}
where the leading thermal exponent $y_{t1}$ is the inverse of the correlation-length exponent, $y_{t1}=1/\nu$, and the leading magnetic exponent $y_{h1}$ corresponds to the fractal dimension $d_f$ of the FK clusters, $y_{h1}=d_f$.
The corresponding exponents for the subleading fields are given by
\begin{align}
y_{t2} & = 4-\frac{4}{g}\,,    \label{eq-yt2} \\ 
y_{h2} & = \frac{(2g+5)(2g-1)}{8g}\,.    \label{eq-yh2}
\end{align}
These subleading fields govern the convergence of corrections to scaling.

\begin{figure}
\centering
\includegraphics[width=\columnwidth]{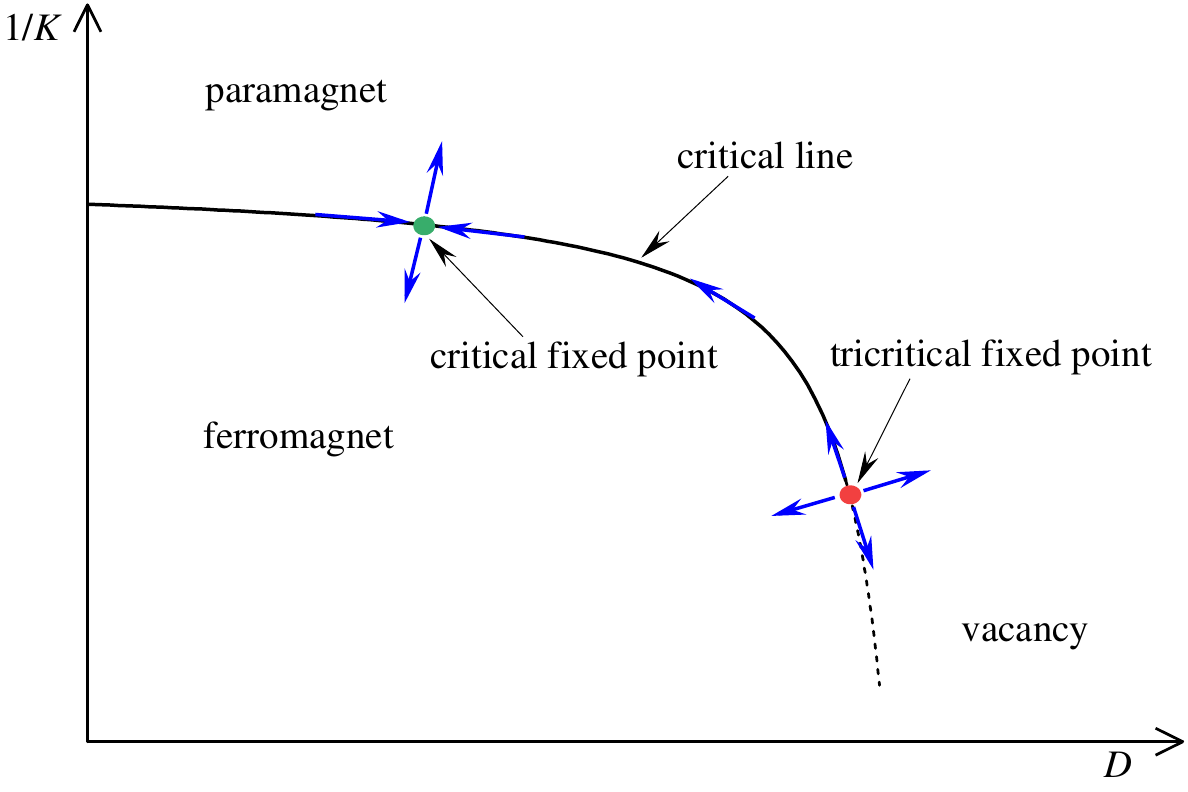}
\caption{(Color online) Schematic phase diagram of the dilute Potts model in the $(D,1/K)$ plane. The solid line denotes the critical line separating the ferromagnetic and paramagnetic phases, while the dashed line indicates the first-order transition. The blue arrows represent the directions of the RG flow. The green dot marks the stable critical fixed point that governs the critical line, and the red dot marks the tricritical fixed point, which is unstable.}   \label{fig1}
\end{figure}

An important extension of the Potts model is the \emph{dilute Potts model}~\cite{Nienhuis_79,Nienhuis_80,Wu_82}, in which lattice sites can be vacant. Its Hamiltonian is given by
\begin{equation}
\mathcal{H}/k_{\rm B}T = -K \sum_{\langle ij \rangle} t_i t_j \delta_{\sigma_i,\sigma_j} - D \sum_i (1-t_i)\,,
\end{equation}
where $t_i = 1$ or $0$ indicates whether site $i$ is occupied or vacant, and $D$ is the reduced chemical potential of the vacancy.

In the limit $D\to -\infty$, all sites are occupied and the model reduces to the standard Potts model. In this pure system, a phase transition from a paramagnetic to a ferromagnetic phase occurs at a critical coupling $K_c$ as $K$ increases (i.e., temperature decreases). When $D$ increases, vacancies appear. The presence of these vacancies shifts the critical coupling $K_c$ to higher values. Along the critical line $(K_c, D_c)$, the universality class of the continuous transition remains that of the standard Potts model up to a tricritical point $(K_t, D_t)$. As illustrated schematically in Fig.~\ref{fig1}, the renormalization-group (RG) flow along this critical line is toward a stable fixed point, where the amplitude of the leading irrelevant thermal scaling field vanishes. The tricritical point $(K_t, D_t)$ corresponds to a distinct, unstable RG fixed point characterized by a different set of critical exponents. The formulas Eqs.~\eqref{eq-yt1}--\eqref{eq-yh2} can be extended to this tricritical point, but with $g\in[1,3/2]$ for $Q\in[4,0]$. For $D>D_t$, vacancies dominate, driving the phase transition first-order, which separates the ferromagnetic and vacancy-dominated phases.

\begin{figure}
\centering
\includegraphics[width=\columnwidth]{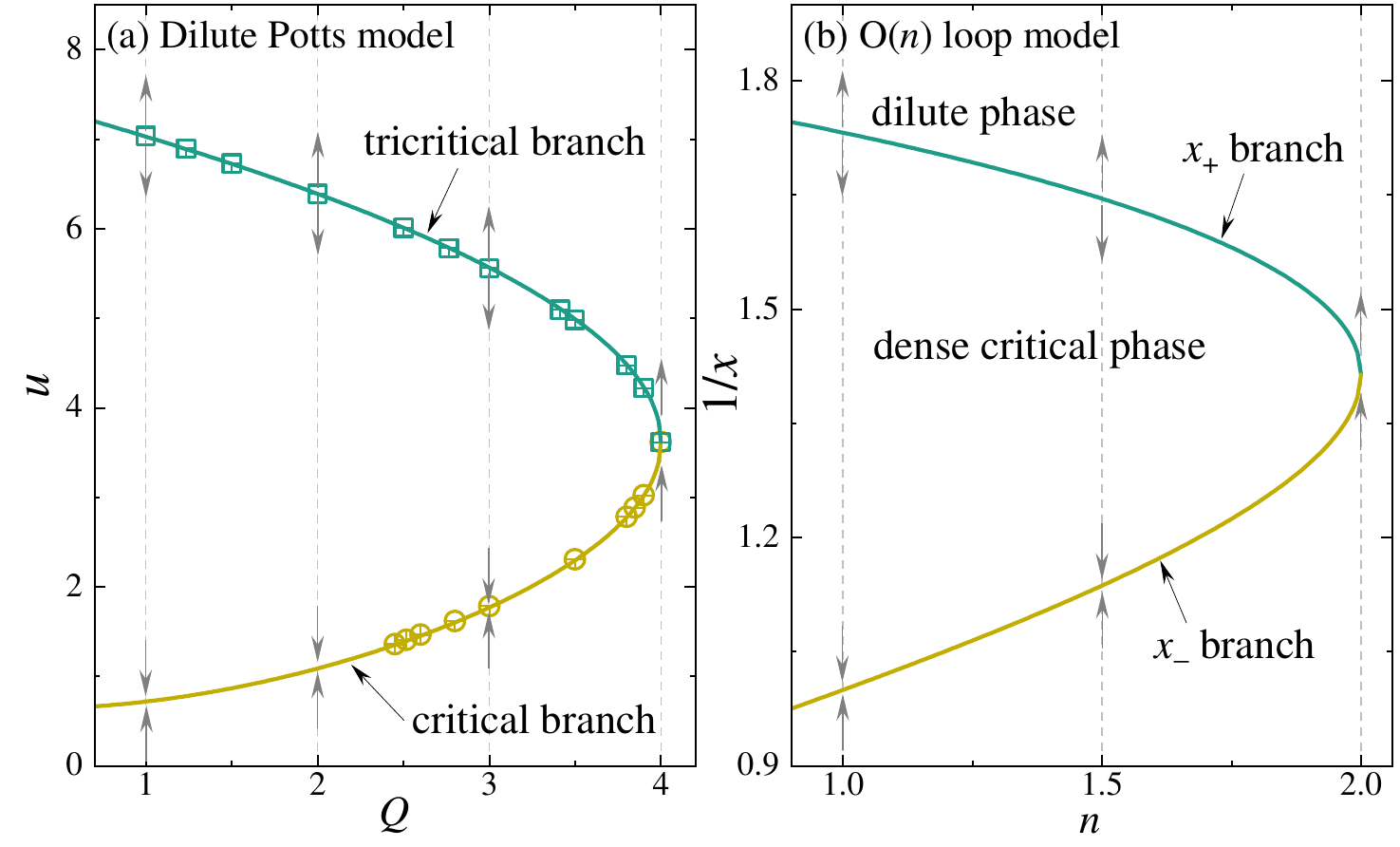}
\caption{(Color online) Correspondence between the phase diagrams of the square-lattice dilute Potts model and the hexagonal-lattice O$(n)$ loop model.
(a) For the dilute $Q$-state Potts model on the square lattice, the quantity $u \equiv e^{D}/(e^{K}-1)$ is calculated from known critical parameters $(K_c, D_c)$ (yellow) and tricritical parameters $(K_t, D_t)$ (green)~\cite{Qian_05}. The yellow branch represents stable fixed points (critical), while the green branch represents unstable fixed points (tricritical). Solid lines are guides to the eye.
(b) For the O$(n)$ loop model on the hexagonal lattice, the analytical branches $x_\pm(n)$ from Eq.~\eqref{eq-xpm} are shown. A continuous phase transition occurs at $x_+$ from a dilute loop phase to a dense, critical phase, whose universality is governed by the stable fixed points at $x_-$.
The mapping $Q = n^2$ establishes the correspondence: the stable critical branch of the Potts model corresponds to the $x_-$ branch, while the unstable tricritical branch corresponds to the $x_+$ critical line.}  \label{fig2}
\end{figure}

To better illustrate how the universality evolves with $Q$, we introduce a ratio~\cite{Qian_05}
\begin{equation}
u \equiv \frac{e^{D}}{e^{K}-1} \,.
\end{equation}
Using the critical $(K_c, D_c)$ and tricritical points $(K_t, D_t)$ for the 2D dilute Potts model (taken from Ref.~\cite{Qian_05}), we identify two smooth branches in the $(Q,u)$ plane, as shown in Fig.~\ref{fig2}(a). The branch formed by the points $(K_c, D_c)$ represents a family of stable fixed points, each corresponding to the critical behavior for a given value of $Q$. The branch formed by $(K_t, D_t)$ corresponds to a family of unstable fixed points that define the tricritical universality classes. From a RG perspective, the distinction between these branches is clear: Along the critical line, the leading thermal scaling field [associated with the energy operator, Eq.~\eqref{eq-yt1}] is relevant, while the subleading thermal field [the dilution operator, Eq.~\eqref{eq-yt2}] is irrelevant. In contrast, at the tricritical fixed point, the dilution operator itself becomes relevant.

A similar two-branch structure is also observed in the phase diagram of the O$(n)$ loop model~\cite{Nienhuis_82,Batchelor_89,Peled_19,Duminil-Copin_21}, as shown in Fig.~\ref{fig2}(b). This model provides a geometrical formulation in terms of nonintersecting loops, characterized by a loop fugacity $n$ and a bond weight $x$. Its partition function is given by
\begin{equation}
\mathcal{Z}_{\text{loop}} = \sum_{\{\mathcal{C}\}} x^{\mathcal{L(C)}} n^{\mathcal{N(C)}}\,,
\end{equation}
where $\mathcal{L(C)}$ and $\mathcal{N(C)}$ denote, respectively, the total loop length and the number of loops in the loop configuration $\mathcal{C}$. On the hexagonal lattice, the two branches shown in Fig.~\ref{fig2}(b) can be obtained analytically as~\cite{Nienhuis_82,Baxter_86,Baxter_87}
\begin{equation}
x_\pm = \frac{1}{\sqrt{2 \pm \sqrt{2 - n}}}\,.   \label{eq-xpm}
\end{equation}
These two branches, $x_+$ and $x_-$, separate distinct critical regimes. When $x$ increases, the system undergoes a continuous transition at $x_+$ from a dilute phase of small loops to a dense, critical phase. The universality class of this dense phase is described by the line of fixed points at $x_-$.
The RG analysis of this model shows that the RG trajectories correspond to constant values of $n$. Therefore, for a fixed value of $n\in [0,2]$, the interval $x\in (0,x_+(n))$ corresponds to the dilute phase, while $x \in (x_+(n),\infty)$ corresponds to the dense critical phase.

The O$(n)$ loop model is believed to be equivalent to the $Q$-state Potts model with $Q=n^2$, sharing the same CG coupling $g$ defined in Eq.~\eqref{eq:Q_vs_g} and thus belonging to the same universality class. This conjecture is supported by the fact that both models have the same central charge in the CG theory and share many critical exponents. However, subtle differences exist. For instance, the leading thermal eigenvalue $y_{t1}$ is not identical to the leading thermal eigenvalue of the Potts model, but rather to its subleading one $y_{t2}$ \cite{Nienhuis_82}. The proposed equivalence specifically implies that the FK clusters of the critical and tricritical Potts models have the same critical behavior as the domains enclosed by the loops in the O$(n)$ model. This hypothesis has been supported by both analytical~\cite{Nienhuis_87} and numerical~\cite{Fang_22,Xu_25,Xu_25a} studies. In particular, Ref.~\cite{Fang_22} provided numerical evidence that the backbone and shortest-path exponents coincide in both models.

A central aspect in the study of these models is the analysis of their geometric structures and the statistical correlations they induce, which reveal the underlying critical behavior. A fundamental example is the two-point connectivity, measuring the probability that two distant sites, located at positions $\bm{x}_1$ and $\bm{x}_2$, belong to the same FK cluster. At criticality, it decays algebraically with distance as
\begin{equation}
P_2^\text{FK}(\bm{x}_1,\bm{x}_2) \sim \frac{A_\text{FK}}{|\bm{x}_1 - \bm{x}_2|^{2\Delta_\text{FK}}}\,,  \label{P2_FK}
\end{equation}
where $\Delta_\text{FK}$ is the scaling dimension associated with the FK clusters, and $A_{\rm FK}$ is the corresponding amplitude. Within the CG theory~\cite{Nienhuis_82,Nienhuis_84,Nienhuis_87}, $\Delta_\text{FK}$ can be exactly determined in 2D and is related to the corresponding fractal dimension via $d_f = y_{h1} = 2 - \Delta_\text{FK}$, where $y_{h1}$ is given by Eq.~\eqref{eq-yh1}.

Beyond the FK clusters themselves, the geometric representation also enables one to investigate more intricate substructures within clusters, such as the so-called \emph{backbone}, which captures the biconnected skeleton after removing all dangling ends and bridges~\cite{Fang_22,Deng_04}. Analogous to the FK clusters, the two-point connectivity function of the backbone at large distances behaves as
\begin{equation}
P_2^\text{BB}(\bm{x}_1,\bm{x}_2) \;\sim\; \frac{ A_\text{BB} }{ |\bm{x}_1-\bm{x}_2|^{2\Delta_\text{BB}} }\,, \label{P2_BB}
\end{equation}
where $\Delta_\text{BB}$ is the scaling dimension associated with the backbone, and $A_\text{BB}$ is the corresponding amplitude. The exactly value of $\Delta_\text{BB}$ was recently determined for the $Q$-state Potts model using the conformal loop ensemble~\cite{Nolin_23,Nolin2025}. Unlike previously known exactly solved critical exponents, the backbone exponent is a transcendental number rather than a rational number. 
For 2D percolation, it is $\Delta_\text{BB} = 0.3566668\ldots$. 
This value is well consistent with the previously reported Monte Carlo estimates of $d_{\rm B}=1.64339(5)$~\cite{Fang_22} via the relation $\Delta_\text{BB} = 2- d_{\rm B}$.

A natural extension is to consider the corresponding three-point functions, $P_3^\text{FK}(\bm{x}_1,\bm{x}_2,\bm{x}_3)$ and $P_3^\text{BB}(\bm{x}_1,\bm{x}_2,\bm{x}_3)$, which represent the probability that the three sites belong to the same FK or backbone cluster, respectively. CFT predicts that~\cite{DiFrancesco_97}, when the three points are far apart, these functions decay as
\begin{align}
P_3^\text{FK}(\bm{x}_1,\bm{x}_2,\bm{x}_3) &\sim \frac{B_\text{FK}}
{r_{12}^{\Delta_\text{FK}} \, r_{13}^{\Delta_\text{FK}} \,
 r_{23}^{\Delta_\text{FK}} }\,, \label{P3_FK} \\
P_3^\text{BB}(\bm{x}_1,\bm{x}_2,\bm{x}_3) &\sim  \frac{B_\text{BB}}
{r_{12}^{\Delta_\text{BB}} \, r_{13}^{\Delta_\text{BB}} \,
 r_{23}^{\Delta_\text{BB}} } \label{P3_BB}  \,, 
\end{align}
where $r_{ij} = |\bm{x}_i - \bm{x}_j|$, and $B_\text{FK}$ ($B_\text{BB}$) is the amplitude for FK (backbone) clusters.

In CFT, the two- and three-point correlation functions \eqref{P2_FK}--\eqref{P3_BB} are often expressed as expectation values of certain fields: $P_2^j(\bm{x}_1,\bm{x}_2) = \langle \Phi_j(\bm{x}_1) \Phi_j(\bm{x}_2) \rangle$ and $P_3^j(\bm{x}_1,\bm{x}_2,\bm{x}_3) = \langle \Phi_j(\bm{x}_1) \Phi_j(\bm{x}_2) \Phi_j(\bm{x}_3) \rangle$, with $j \in \{\text{FK},\text{BB}\}$. By choosing a suitable normalization of the field $\Phi_j$, the constants in the two-point functions can be set to $A_j = 1$, and the constants $B_j$ in the three-point functions then define the so-called three-point structure constants $R_\text{FK}$ and $R_\text{BB}$. These constants can also characterize the universality class of the corresponding model.

On the lattice, the normalization of the fields used in CFT is not directly possible. To extract the universal three-point structure constants $R_j$, one considers the ratio
\begin{equation}
R_j(\bm{x}_1,\bm{x}_2,\bm{x}_3)
\;=\; \frac{ P_3^j(\bm{x}_1,\bm{x}_2,\bm{x}_3) }
{\sqrt{P^j_2(\bm{x}_1,\bm{x}_2) P^j_2(\bm{x}_1,\bm{x}_3) P^j_2(\bm{x}_2,\bm{x}_3)}}\,, \label{def_Rxyz}
\end{equation}
where the points $\bm{x}_1,\bm{x}_2,\bm{x}_3$ are chosen to be mutually far apart. This ratio eliminates nonuniversal metric factors, so that $R_j(\bm{x}_1,\bm{x}_2,\bm{x}_3)$ converges to the universal constant $R_j$ in the thermodynamic limit. For finite lattices, however, some residual dependence on the specific choice of points is expected.

The universal three-point amplitude ratio $R_\text{FK}$ for the FK clusters in the 2D $Q$-state Potts model has been exactly determined from CFT~\cite{Delfino_11,Picco_13}. For percolation ($Q=1$ in the critical regime), numerical results consistent with the exact solution $R_{\text{FK}} = 1.02201\ldots$ were reported both before~\cite{Simmons2009} and after~\cite{Ziff_11b} this exact result was established. In addition to the FK clusters, an analogous ratio can be defined for spin clusters, $R_\text{spin}$. Numerical work by Ref.~\cite{Delfino_13} provided values for $R_\text{spin}$ for selective $Q$, and its exact solution for general $Q$ was recently obtained in Ref.~\cite{Cai2025}.

In this work, we perform large-scale Monte Carlo simulations to determine the universal three-point amplitude ratio $R_\text{BB}$ associated with the backbone structure in both critical and tricritical Potts models. Since for $Q$ close to $4$, direct simulations of the Potts model suffer from severe critical slowing down and slowly converging finite-size corrections, we instead adopt the O$(n)$ loop model on the hexagonal lattice~\cite{Nienhuis_82,Batchelor_89,Peled_19,Duminil-Copin_21}, for which highly efficient algorithms are available~\cite{Fang_22}. We measure the universal ratio $R_\text{BB}$ and, for reference, $R_\text{FK}$ whose exact values are known from CFT~\cite{Delfino_11,Picco_13}.

\begin{figure}
\centering
\includegraphics[width=\columnwidth]{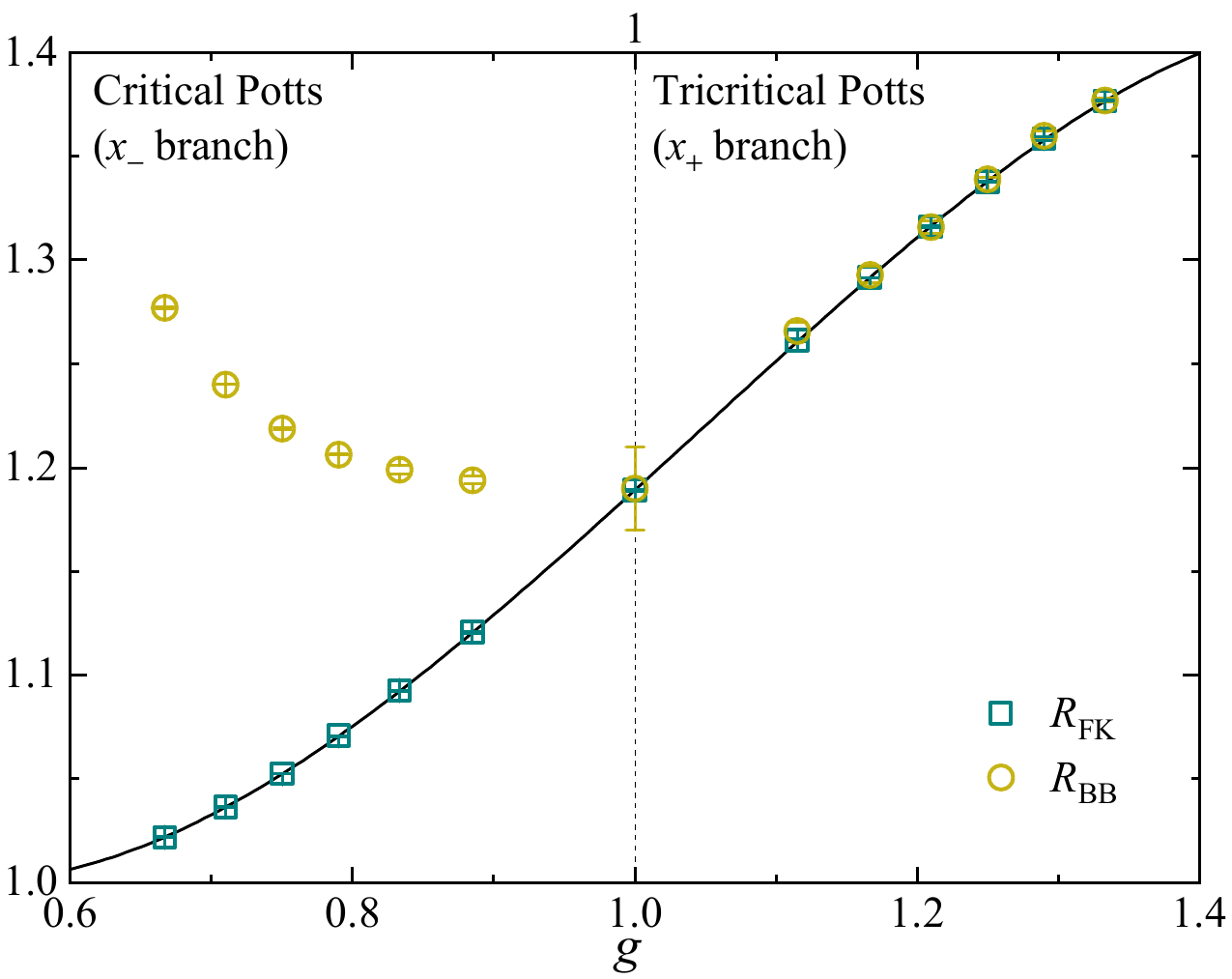}
\caption{(Color online) Three-point structure constant of the Potts model as a function of the Coulomb-gas coupling $g$. The solid line represents the exact result for $R_\text{FK}$ (see the Appendix for calculation details)~\cite{Delfino_11,Picco_13}. The Monte Carlo data (Table~\ref{table:R}) are shown as green squares for $R_\text{FK}$ and yellow circles for $R_\text{BB}$.}  \label{fig3}
\end{figure}

Our numerical results reveal that $R_\text{BB}$ is systematically larger than $R_\text{FK}$ along the critical branch, but the two quantities coincide within error bars in the tricritical regime. This mirrors the correspondence previously observed in two-point functions, where the backbone fractal dimension $d_{\rm B}$ equals the FK cluster dimension $d_f$ at tricriticality~\cite{Deng_04,Fang_22}. Together, these findings provide strong evidence that the backbone and FK clusters belong to distinct universality classes along the critical branch, but merge into a common universality at the tricritical point. A summary of our numerical estimates and theoretical predictions for $R_\text{FK}$ is presented in Table~\ref{table:R} and Fig.~\ref{fig3}.

\begin{table}
\centering
\caption{Three-point structure constants $R_\text{FK}$ and $R_\text{BB}$ for the $Q$-state critical and tricritical Potts models. The data points are listed in increasing order of the Coulomb-gas coupling $g$. The interval $g\in [2/3,1]$ corresponds to the critical regime, while the interval $g\in [1,4/3]$, to the tricritical regime.} \label{table:R}
\begin{ruledtabular}
\begin{tabular}{lllll}
\multicolumn{1}{c}{$g$} & \multicolumn{1}{c}{$Q$}  &  \multicolumn{1}{c}{$R_\text{FK}$(Theory)} &  \multicolumn{1}{c}{$R_\text{FK}$} & \multicolumn{1}{c}{$R_\text{BB}$} \\
\hline
2/3    & 1   & 1.02201\ldots &  1.02199(5) & 1.2771(4) \\
0.7098 & 1.5 & 1.03639\ldots &  1.03638(8) & 1.2401(1) \\
3/4    & 2   & 1.05245\ldots &  1.05252(9) & 1.2188(6) \\
0.7902 & 2.5 & 1.07071\ldots &  1.0707(1)  & 1.2063(2) \\
5/6    & 3   & 1.09236\ldots &  1.0926(3)  & 1.199(2)  \\
0.8850 & 3.5 & 1.12056\ldots &  1.1206(5)  & 1.194(2)  \\
\hline
1      & 4   & 1.18920\ldots &  1.1891(2)  & 1.19(2)   \\
\hline
1.1150 & 3.5 & 1.26051\ldots &  1.2614(6)  & 1.266(4)  \\
7/6    & 3   & 1.29156\ldots &  1.2917(1)  & 1.293(4)  \\
1.2098 & 2.5 & 1.31629\ldots &  1.3160(6)  & 1.316(3)  \\
5/4    & 2   & 1.33794\ldots &  1.3378(2)  & 1.339(1)  \\
1.2902 & 1.5 & 1.35784\ldots &  1.3586(9)  & 1.360(2)  \\
4/3    & 1   & 1.37673\ldots &  1.3768(3)  & 1.3770(8) \\
\end{tabular}
\end{ruledtabular}
\end{table}

The remainder of the paper is organized as follows. In Sec.~\ref{sec:MC}, we describe the numerical methods based on the O$(n)$ loop model. Section~\ref{sec:data} details the analysis of the Monte Carlo data, and Sec.~\ref{sec:res} discusses the numerical results and their implications.
Finally, in the Appendix, we show the details of the exact computation of $R_\text{FK}$.

\section{Monte Carlo simulations}  \label{sec:MC}

In this section, we describe the Monte Carlo simulations used to generate FK and backbone clusters in both the critical and tricritical Potts models. To this end, we employ the O$(n)$ loop model on the hexagonal lattice, which is equivalent to a generalized Ising model on its dual triangular lattice. Section~\ref{sec:MC:algo} outlines the simulation algorithm, while Sec.~\ref{sec:MC:obs} details the measurement of the two- and three-point correlation functions.

\subsection{Monte Carlo algorithm} \label{sec:MC:algo}

As explained in the Introduction, we adopt the O$(n)$ loop model on the hexagonal lattice rather than directly simulating the corresponding $Q$-state Potts models at their critical or tricritical points. The loops on the hexagonal lattice correspond to the boundaries of Ising domains on the dual triangular lattice, establishing a one-to-two correspondence between loop configurations and dual Ising-spin configurations. The dual coupling strength satisfies $2K^* = -\ln(x)$, being ferromagnetic for $x < 1$ and antiferromagnetic for $x > 1$, while $n$ serves as the fugacity associated with each Ising domain. The partition function of this generalized Ising model on the triangular lattice can be written as
\begin{equation}
\mathcal{Z}_{\rm GIsing}=\sum_{\{\sigma\}} n^{N_d} \prod_{\langle i,j\rangle}e^{K^*\sigma_i\sigma_j},
\end{equation}
where $N_d$ is the number of dual Ising domains.

Following the algorithm in Ref.~\cite{Fang_22}, the update procedure of this generalized Ising model on the triangular lattice proceeds iteratively as follows:
\begin{enumerate}
  \item For each Ising domain, all sites are independently set to be ``active'' with probability $1/n$, and ``inactive'' otherwise. We restrict to $n \in [1,2]$.
  \item Bonds between neighboring sites are then occupied according to:
  \begin{enumerate}
    \item If both sites are active and have the same spin value, occupy the bond with probability $p = 1 - x_\pm$, where $x_\pm$ is given by Eq.~\eqref{eq-xpm}.
    \item If both sites are active but have opposite spins, leave the bond empty.
    \item If at least one of the adjacent sites is inactive, occupy the bond with probability one. 
  \end{enumerate}
  \item New Ising domains are obtained by flipping the spins of all sites in each connected component of occupied bonds, with probability $1/2$. Note that sites within a connected component may carry different spin values [because of point 2(c) above].
\end{enumerate}

The resulting Ising domains correspond to FK clusters of the Potts model, with $x_-$ and $x_+$ representing the critical and tricritical regimes, respectively. The associated backbone structure is extracted by scanning each FK cluster using a depth-first search. The two- and three-point correlation functions are then sampled as described in Sec.~\ref{sec:MC:obs}.

It is worth noting that the FK clusters generated by this algorithm correspond to site configurations (Ising domains), whereas those obtained by the Swendsen--Wang cluster algorithm~\cite{Swendsen_87} correspond to bond configurations. A special case occurs at $n = \sqrt{Q} = 1$, where the Swendsen--Wang cluster algorithm reduces to standard bond percolation, while our algorithm yields site percolation. Although both types of clusters share identical large-scale geometric properties at criticality, they differ in how correlation functions are computed. For bond percolation, the correlation function is obtained by scanning all clusters, whereas for site percolation, only clusters formed by occupied sites are considered.

Because of spin symmetry in this generalized Ising model, FK clusters corresponding to the two spin values exhibit identical statistics. However, similar to site percolation, when computing correlation functions for $Q > 1$, only clusters of one spin species are considered. Including both spin species would make the ratios $R_j$ defined in Eq.~\eqref{def_Rxyz} smaller by a factor of $1/\sqrt{2}$ than those defined for standard (bond) FK clusters, due to double counting. This subtlety was also discussed in the Supplementary Material of Ref.~\cite{Ikhlef_16}. In our simulations, observables are measured separately for both spin species, and the results are averaged.

The equivalence between the O$(n)$ loop model on the hexagonal lattice and the generalized Ising model on its dual triangular lattice holds exactly for planar lattices. In practice, however, we simulate finite systems with periodic boundary conditions, which break planarity. Such boundary effects are expected to have negligible influence on bulk critical behavior, and indeed no observable deviation has been reported in previous studies~\cite{Fang_22,Xu_25,Xu_25a}.

We performed simulations of the O$(n = \sqrt{Q})$ loop model for $Q = 1, 1.5, 2, 2.5, 3, 3.5, 4$, on both branches $x_-$ and $x_+$. For each $n$, we used triangular lattices with linear sizes $L = 32, 64, 128, 256, 512, 1024, 2048, 4096, 8192$. The number of statistically independent samples ranged from $10^8$ for the smallest systems down to $10^3$ for the largest ones.

\subsection{Observables} \label{sec:MC:obs}

As discussed in the Introduction, the primary quantity of interest is the ratio $R_j(\bm{x}_1,\bm{x}_2,\bm{x}_3)$ defined in Eq.~\eqref{def_Rxyz}. To compute this ratio efficiently, we select the triplet $(\bm{x}_1,\bm{x}_2,\bm{x}_3)$ such that they form an equilateral triangle with side length $r \equiv |\bm{x}_1-\bm{x}_2| = |\bm{x}_1-\bm{x}_3| = |\bm{x}_2-\bm{x}_3|$ (recall that the model is defined on a triangular lattice; see Sec.~\ref{sec:MC:algo}). For each site $\bm{x}_1$, several pairs $(\bm{x}_2,\bm{x}_3)$ satisfy this condition. In our simulations, we select one equilateral triangle such that $\bm{x}_2$ shares the same horizontal coordinate as $\bm{x}_1$, while $\bm{x}_3$ lies above them with a larger vertical coordinate.

Traversing $\bm{x}_1$ over all lattice sites in a given configuration, we record the number of occurrences $N^j_{12}$ that sites $\bm{x}_1$ and $\bm{x}_2$ belong to the same cluster of type $j$ (the other pairs are treated analogously). When checking whether two sites are connected, the connectivity of the third site is not considered. Similarly, $N^j_{123}$ counts the occurrences where all three sites belong to the same cluster.

For $M$ independent realizations, the two-point (e.g., $\bm{x}_1$ and $\bm{x}_2$) and three-point correlation functions are estimated as
\begin{align}
P^j_2(\bm{x}_1,\bm{x}_2) &= \frac{1}{M V} \sum_{i=1}^{M} N^{j}_{1,2}(i)\,,   \\
P^j_3(\bm{x}_1,\bm{x}_2,\bm{x}_3) &= \frac{1}{M V} \sum_{i=1}^{M} N^{j}_{1,2,3}(i)\,, \label{def_Ps}
\end{align}
where $V$ is the total number of lattice sites, and the sum runs over all realizations. Finally, substituting these correlation functions into Eq.~\eqref{def_Rxyz}, the ratio $R_j(\bm{x}_1,\bm{x}_2,\bm{x}_3)$ can be obtained as a function of $r$.

\section{Data analysis}    \label{sec:data}

In this section we describe how the three-point structure constants $R_\text{FK}$ and $R_\text{BB}$ were obtained from our Monte Carlo data,
as reported in Table~\ref{table:R}. The analysis focuses on integer values $Q=1, 2, 3, 4$ on both the critical ($x_-$) and tricritical ($x_+$) regimes. Results for noninteger $Q$ ($=1.5, 2.5, 3.5$) follow qualitatively the same behavior and are omitted for brevity.

\subsection{General procedure}

\begin{figure*}
\centering
\includegraphics[width=2.0\columnwidth]{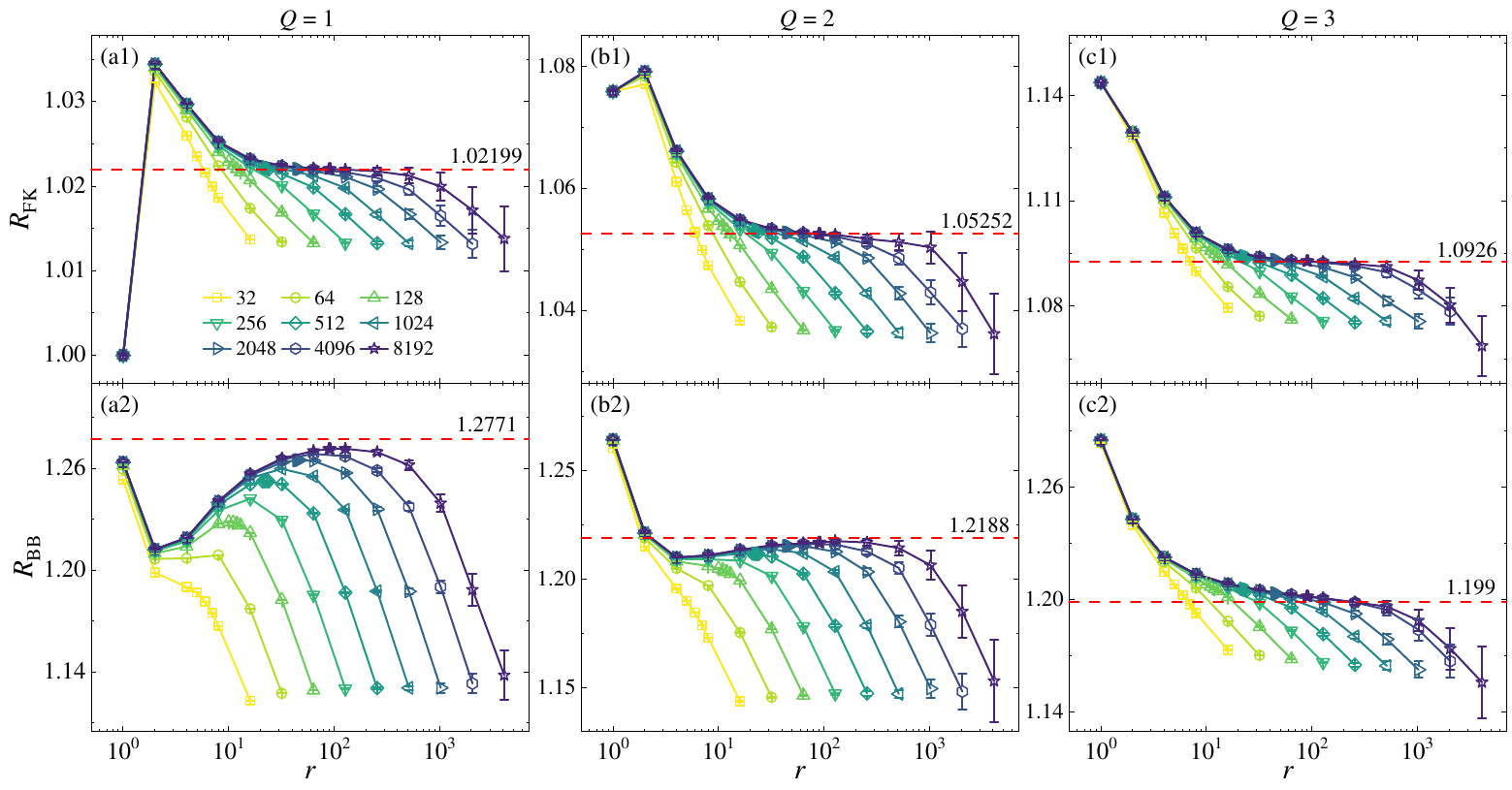}
\caption{(color online) Three-point structure constants $R_\text{FK}$ and $R_\text{BB}$ as functions of $r$ on the $x_-$ branch of the O$(n)$ loop model with $Q=n^2$. Panels (a), (b), and (c) are for $Q=1,2,3$, respectively. Data are shown for linear sizes $L=32,64,\ldots,4096,8192$, with each size represented by a distinct color. Dashed lines indicate the fits reported in Table~\ref{table:R}.}   \label{fig4}
\end{figure*}

\begin{figure*}
\centering
\includegraphics[width=2.0\columnwidth]{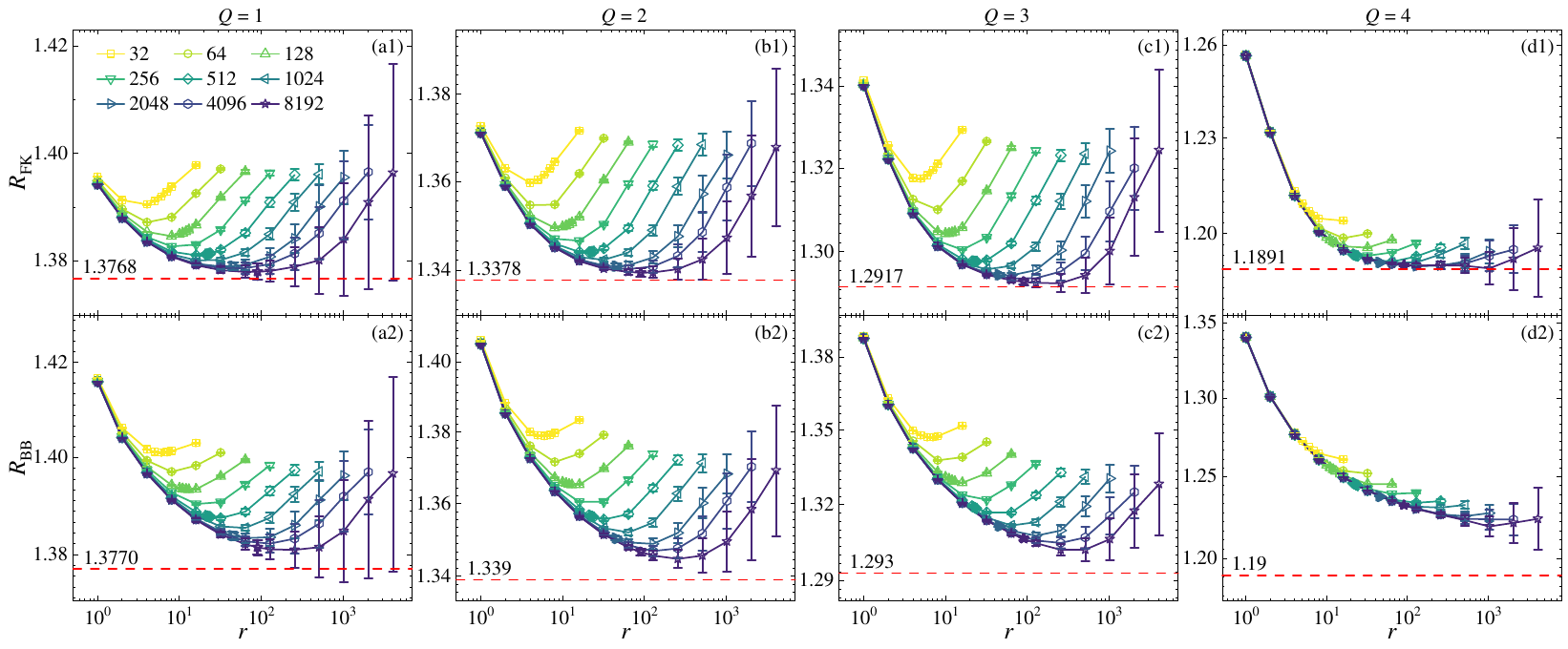}
\caption{Three-point structure constants $R_\text{FK}$ and $R_\text{BB}$ as functions of $r$ on the $x_+$ branch of the O$(n)$ loop model with $Q=n^2$. Panels (a), (b), (c), and (d) are for $Q=1,2,3,4$, respectively. Data are shown for linear sizes $L=32,64,\ldots,4096,8192$, with each size represented by a distinct color. Dashed lines indicate the fits reported in Table~\ref{table:R}.}    \label{fig5}
\end{figure*}

For equilateral triangles of side length $r$, we denote the three-point ratios as $R_j(r)$ with $j\in\{\text{FK},\text{BB}\}$. The raw data show a weak, yet noticeable, dependence on the triangle size $r$ (see Figs.~\ref{fig4} and \ref{fig5}). For the $x_-$ branch, $R_{\rm FK}(r)$ exhibits an inflection point and $R_{\rm BB}(r)$ a local maximum near $r \simeq \sqrt{L/2}$, while for the $x_+$ branch a minimum occurs for both quantities at the same value. The case $Q=4$ is less clear, especially for $R_\text{BB}$, indicating slower convergence.

To extract the values in the thermodynamic limit, we focus on $r = \sqrt{L/2}$ and fit the corresponding data to the ansatz
\begin{equation}
R_j(\sqrt{L/2}) = R_j + L^{-y} \left( a_0 + a_1 L^{-y'} \right)\,,  \label{eq-rjl}
\end{equation}
where $R_j$ is the value in the infinite-volume limit and $y,y'>0$ are correction-to-scaling exponents. For system sizes where $r=\sqrt{L/2}$ is not integer, $R_j(r)$ is estimated via linear interpolation between the two nearest integer $r$ values. To check the robustness of this procedure, we tried a polynomial interpolation based on the four nearest integer values; the differences were negligible.

It is worth noting that fitting all five free parameters in the ansatz \eqref{eq-rjl} (i.e., $R_j$, $y$, $y'$, $a_0$, and $a_1$) simultaneously often leads to unstable results. In practice, we therefore fixed the correction-to-scaling exponent $y'$ to a reasonable value (e.g., $y'=0.25, 0.5, 1, 2$) and performed the fit for the remaining four parameters $\{R_j, y, a_0, a_1\}$.

To assess possible corrections to scaling not captured by the ansatz \eqref{eq-rjl}, we repeated the fits using only data with $L \ge L_\text{min}$ and examined how the estimated parameters vary with $L_\text{min}$. In general, our preferred fit corresponds to the smallest $L_\text{min}$ for which the goodness of fit is reasonable, and for which further increases of $L_\text{min}$ do not reduce $\chi^2$ by significantly more than one unit per degree of freedom. If increasing $L_\text{min}$ produces a monotonic trend, we consider the fit has not converged, even if the $\chi^2$ reduction is moderate. For each fit, we report $\chi^2$ and the number of degrees of freedom (DF). Since $y'$ is fixed and additional correction terms may be present, the statistical error from a single fit may be underestimated. Therefore, the final error quoted in Table~\ref{table:R} also accounts for the observed variation of $R_j$ when $L_\text{min}$ and/or $y'$ are varied, providing a conservative (though subjective) estimate of uncertainty.

\subsection{\texorpdfstring{$x_-$ branch}{xminus branch}}

\begin{figure*}
\centering
\includegraphics[width=2.0\columnwidth]{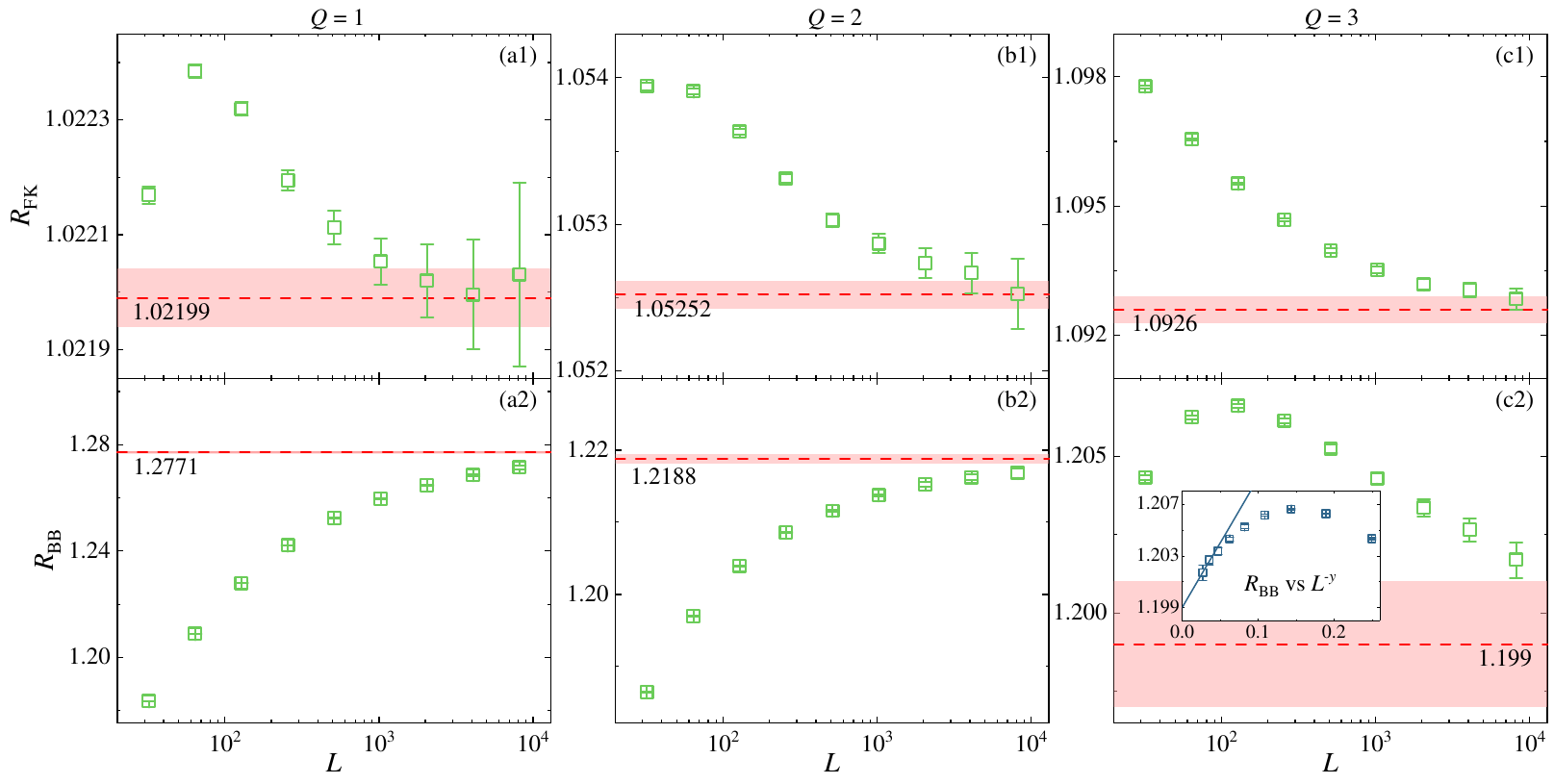}
\caption{Three-point structure functions $R_\text{FK}(\sqrt{L/2})$ and $R_\text{BB}(\sqrt{L/2})$ as a function of the linear size of the system $L$ on the $x_-$ branch. Panels (a), (b), and (c) are for $Q=1,2,3$, respectively. The dashed red lines show the final result reported on Table~\ref{table:R}, and the shaded regions depict the corresponding error bars. The inset of panel (d2) plots $R_\text{BB}(\sqrt{L/2})$ as a function of $L^{-y}$ with $y=0.4$, see the text for a discussion.}
\label{fig6}
\end{figure*}

\begin{table}
\caption{Fits of the data $R_\text{FK}(\sqrt{L/2})$ to the ansatz \eqref{eq-rjl} for $Q=1, 2$, and $3$ on the $x_-$ branch. Entries without errors represent fixed values in the corresponding (biased) fit. The fits with $L_\text{min}$ in boldface correspond to our preferred ones.}
\begin{ruledtabular}
\begin{tabular}{lrlllllr}
\multicolumn{1}{c}{$Q$}  &
\multicolumn{1}{c}{$L_{\text{min}}$}  &
\multicolumn{1}{c}{$R_\text{FK}$}  &
\multicolumn{1}{c}{$y$}  &
\multicolumn{1}{c}{$a_0$}  &
\multicolumn{1}{c}{$y'$}  &
\multicolumn{1}{c}{$a_1$}  &
\multicolumn{1}{c}{$\chi^2/\text{DF}$} \\
\hline
\multirow{4}{*}{$1$}
     &  $32$   & $1.02196(1)$   & $0.88(3)$   & $0.045(6)$    & $0.5$    & $-0.23(3)$    &  $0.45/5$   \\
     &  $64$   & $1.02200(1)$   & $1.07(6)$   & $0.12(3)$     & $0.5$    & $-0.7(2)$     &  $0.11/4$   \\
     &  $32$   & $1.02194(2)$   & $0.69(4)$   & $0.013(2)$    & $1$      & $-0.34(6)$    &  $0.66/5$   \\
     &  $\mathbf{64}$   & $1.02199(1)$   & $0.92(5)$   & $0.038(9)$    & $1$      & $-1.3(4)$     &  $0.10/4$   \\
\hline
\multirow{4}{*}{$2$}
     &  $32$   & $1.05243(5)$   & $0.62(3)$   & $0.035(4)$    & $0.5$    & $-0.13(2)$    &  $0.69/5$  \\
     &  $\mathbf{64}$   & $1.05252(4)$   & $0.73(5)$   & $0.06(2)$     & $0.5$    & $-0.27(9)$    &  $0.29/4$  \\
     &  $32$   & $1.05234(8)$   & $0.48(4)$   & $0.015(2)$    & $1$      & $-0.20(4)$    &  $1.02/5$  \\
     &  $64$   & $1.05249(5)$   & $0.62(5)$   & $0.027(6)$    & $1$      & $-0.6(2)$     &  $0.29/4$  \\
\hline
\multirow{4}{*}{$3$}
     &  $32$   & $1.0923(1)$    & $0.56(5)$   & $0.07(2)$     & $0.25$    & $-0.08(3)$   &  $1.93/5$  \\
     &  $\mathbf{64}$   & $1.0926(1)$    & $0.68(7)$   & $0.15(5)$     & $0.25$    & $-0.2(1)$    &  $0.96/4$  \\
     &  $32$   & $1.0923(1)$    & $0.50(4)$   & $0.041(8)$    & $0.5$     & $-0.06(2)$   &  $2.10/5$  \\
     &  $64$   & $1.0925(1)$    & $0.61(6)$   & $0.07(2)$     & $0.5$     & $-0.2(1)$    &  $1.01/4$  \\
\end{tabular}
\end{ruledtabular}
\label{table:FK_xminus}
\end{table}

\begin{table}
\caption{Fits of the data $R_\text{BB}(\sqrt{L/2})$ to the ansatz \eqref{eq-rjl} for $Q=1, 2$, and $3$ on the $x_-$ branch. Entries without errors represent fixed values in the corresponding (biased) fit. The fits with $L_\text{min}$ in boldface correspond to our preferred ones.}
\begin{ruledtabular}
\begin{tabular}{lrlllllr}
\multicolumn{1}{c}{$Q$}  &
\multicolumn{1}{c}{$L_{\text{min}}$}  &
\multicolumn{1}{c}{$R_\text{BB}$}  &
\multicolumn{1}{c}{$y$}  &
\multicolumn{1}{c}{$a_0$}  &
\multicolumn{1}{c}{$y'$}  &
\multicolumn{1}{c}{$a_1$}  &
\multicolumn{1}{c}{$\chi^2/\text{DF}$} \\
\hline
\multirow{4}{*}{$1$}
     &  $32$    & $1.2770(2)$    & $0.552(4)$   & $-0.91(2)$    & $0.25$  & $-0.65(3)$    &  $0.74/5$  \\
     &  $64$    & $1.2768(3)$    & $0.56(1)$    & $-0.94(6)$    & $0.25$  & $-0.7(1)$     &  $0.68/4$  \\
     &  $32$    & $1.2775(2)$    & $0.513(4)$   & $-0.64(1)$    & $0.5$   & $-0.46(3)$    &  $0.97/5$  \\
     &  $\mathbf{64}$    & $1.2771(3)$    & $0.523(7)$   & $-0.67(3)$    & $0.5$   & $-0.57(8)$    &  $0.63/4$  \\
\hline
\multirow{4}{*}{$2$}
     &  $32$    & $1.2184(2)$    & $0.52(1)$    & $-0.17(1)$    & $0.5$   & $-0.15(2)$   &  $0.35/5$  \\
     &  $\mathbf{64}$    & $1.2188(2)$    & $0.47(3)$    & $-0.12(2)$    & $0.5$   & $-0.24(3)$   &  $0.15/4$  \\
     &  $32$    & $1.2182(1)$    & $0.556(9)$   & $-0.210(7)$   & $1$     & $-0.28(5)$   &  $0.53/5$  \\
     &  $64$    & $1.2185(1)$    & $0.52(1)$    & $-0.18(1)$    & $1$     & $-0.7(1)$    &  $0.15/4$   \\
\hline
\multirow{6}{*}{$3$}
     &  $32$    & $1.1988(2)$    & $0.45(1)$    & $-0.35(1)$     & $0.1$    & $-0.46(2)$   &  $0.32/5$  \\
     &  $64$    & $1.1992(3)$    & $0.47(2)$    & $-0.38(3)$     & $0.1$    & $-0.50(4)$   &  $0.23/4$  \\
     &  $32$    & $1.1985(3)$    & $0.37(1)$    & $-0.118(3)$    & $0.25$   & $-0.231(9)$  &  $0.35/5$  \\
     &  $\mathbf{64}$    & $1.1990(4)$    & $0.40(2)$    & $-0.13(1)$     & $0.25$   & $-0.26(2)$   &  $0.23/4$  \\
     &  $32$    & $1.1973(5)$    & $0.26(1)$    & $-0.049(1)$    & $0.5$    & $-0.179(9)$  &  $0.52/5$  \\
     &  $64$    & $1.1982(5)$    & $0.30(2)$    & $-0.055(3)$    & $0.5$    & $-0.22(2)$   &  $0.25/4$
\end{tabular}
\end{ruledtabular}
\label{table:BB_xminus}
\end{table}

In this section we focus on the numerical fits for the critical $Q$-state Potts model, i.e., the O$(n)$ loop model with $n=\sqrt{Q}$ on the $x_-$ branch. Figure~\ref{fig6} shows the data $R_\text{FK}(\sqrt{L/2})$ and $R_\text{BB}(\sqrt{L/2})$ for $Q=1,2,3$. In general, $R_\text{FK}(\sqrt{L/2})$ converges faster, allowing precise fits with small error bars. The data for $R_\text{BB}(\sqrt{L/2})$ are noisier, particularly for $Q=3$, but fits using Eq.~\eqref{eq-rjl} still provide reliable estimates (see Table~\ref{table:R}).

We now discuss the fits for $R_\text{FK}(\sqrt{L/2})$ with different $Q$. The results are summarized in Table~\ref{table:FK_xminus}, where for brevity we only show the cases $Q=1, 2$, and $3$ for $L_\text{min}=32,64$. Two values of $y'$ were tested for $Q<3$ and three for $Q\geq 3$ (only two are shown in the table). In all cases, the fits for different $y'$ were very similar, indicating that $y'$ has little effect on $R_\text{FK}$.

The fits for $R_\text{BB}(\sqrt{L/2})$ with $Q=1,2,3$ are presented in Table~\ref{table:BB_xminus}. For $Q<3$, the ansatz \eqref{eq-rjl} describes the data well, and the results are insensitive to $y'$. Error bars, however, are roughly an order of magnitude larger than for $R_\text{FK}(\sqrt{L/2})$. For $Q=3$, the data are not yet close to their asymptotic values [see Fig.~\ref{fig6}(c2)], so the fits depend more strongly on $y'$. For $y'=0.25$ and $Q=3$, the preferred fit with $L_\text{min}=64$ gives $R_\text{BB}=1.1990(4)$ and $y=0.40(2)$. The inset of Fig.~\ref{fig6}(c2) shows $R_{\text{BB}}(\sqrt{L/2})$ vs $L^{-y}$ with $y=0.40$, where the straight line depicts the preferred fit, indicating convergence to the infinite-volume limit.

\subsection{\texorpdfstring{$x_+$ branch}{xplus branch}} \label{sec:tricrit}

\begin{figure*}
\centering
\includegraphics[width=2.0\columnwidth]{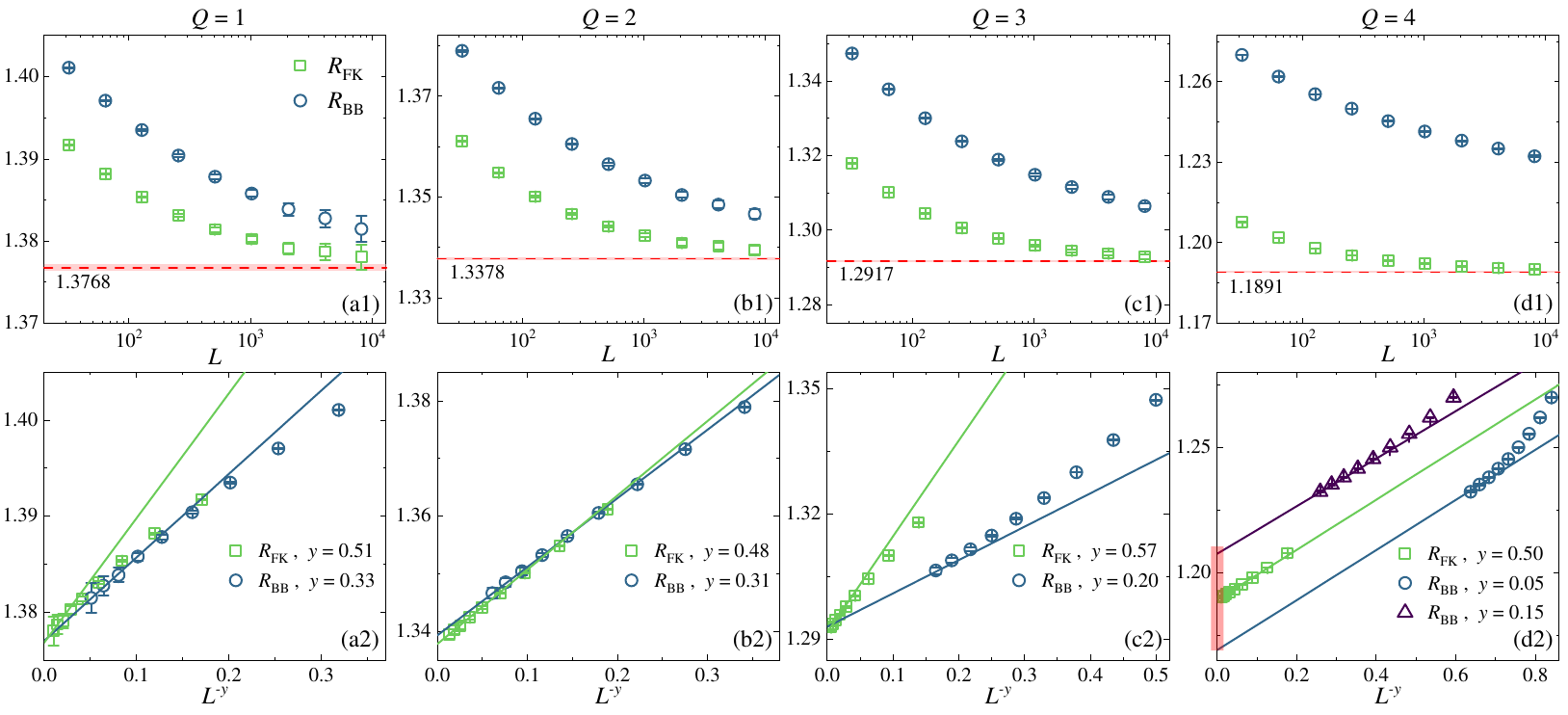}
\caption{(color online) Three-point structure functions $R_\text{FK}(\sqrt{L/2})$ and $R_\text{BB}(\sqrt{L/2})$ on the $x_+$ branch. Top panels (a1)--(d1) show the behavior of $R_\text{FK}(\sqrt{L/2})$ and $R_\text{BB}(\sqrt{L/2})$ as a function of the linear size $L$ for $Q=1$ (a1), $Q=2$ (b1), $Q=3$ (c1), and $Q=4$ (d1). The dashed lines represent the values resulting from the fits of $R_\text{FK}$ to the ansatz \eqref{eq-rjl}, and the shaded regions correspond to the error bars (see Table~\ref{table:R}). Bottom panels (a2)--(d2) show the same quantities as a function of $L^{-y}$, where the value of $y$ has been taken from the corresponding preferred fit of the data to the ansatz \eqref{eq-rjl}. The lines show the leading term of the full ansatz \eqref{eq-rjl}. See Sec.~\ref{sec:potts4} for a discussion of panel (d2) concerning $Q=4$.}
\label{fig7}
\end{figure*}

\begin{table}
\caption{Fits of the data $R_\text{FK}(\sqrt{L/2})$ to the ansatz \eqref{eq-rjl} for $Q=1, 2$, and $3$ on the $x_+$ branch. Entries without errors represent fixed values in the corresponding (biased) fit. The fits with $L_\text{min}$ in boldface correspond to our preferred ones.}
\begin{ruledtabular}
\begin{tabular}{lrlllllr}
\multicolumn{1}{c}{$Q$}  &
\multicolumn{1}{c}{$L_{\text{min}}$}  &
\multicolumn{1}{c}{$R_\text{FK}$}  &
\multicolumn{1}{c}{$y$}  &
\multicolumn{1}{c}{$a_0$}  &
\multicolumn{1}{c}{$y'$}  &
\multicolumn{1}{c}{$a_1$}  &
\multicolumn{1}{c}{$\chi^2/\text{DF}$} \\
\hline
\multirow{4}{*}{$1$}
     &  $\mathbf{32}$    & $1.3768(2)$    & $0.51(2)$    & $0.13(2)$     & $0.25$  & $-0.11(2)$   &  $0.12/5$  \\
     &  $64$    & $1.3770(4)$    & $0.53(6)$    & $0.15(5)$     & $0.25$  & $-0.14(9)$   &  $0.11/4$  \\
     &  $32$    & $1.3767(2)$    & $0.46(4)$    & $0.087(7)$    & $0.5$   & $-0.07(2)$   &  $0.12/5$  \\
     &  $64$    & $1.3769(4)$    & $0.49(5)$    & $0.10(2)$     & $0.5$   & $-0.11(7)$   &  $0.11/4$  \\
\hline
\multirow{4}{*}{$2$}
     &  $32$    & $1.3380(2)$    & $0.50(1)$    & $0.147(9)$    & $0.5$   & $-0.08(2)$   &  $0.11/5$  \\
     &  $64$    & $1.3379(3)$    & $0.49(3)$    & $0.14(2)$     & $0.5$   & $-0.06(7)$   &  $0.11/4$  \\
     &  $\mathbf{32}$    & $1.3378(1)$    & $0.483(8)$   & $0.129(4)$    & $1$     & $-0.14(4)$   &  $0.11/5$  \\
     &  $64$    & $1.3378(2)$    & $0.48(2)$    & $0.13(1)$     & $1$     & $-0.1(2)$    &  $0.11/4$  \\
\hline
\multirow{4}{*}{$3$}
     &  $\mathbf{32}$    & $1.2917(1)$    & $0.57(1)$    & $0.23(2)$     & $0.25$   & $-0.10(3)$   &  $0.08/5$  \\
     &  $64$    & $1.2916(2)$    & $0.54(3)$    & $0.19(4)$     & $0.25$   & $-0.04(6)$   &  $0.07/4$  \\
     &  $32$    & $1.29166(9)$   & $0.549(8)$   & $0.189(8)$    & $0.5$     & $-0.07(2)$   &  $0.08/5$  \\
     &  $64$    & $1.2916(1)$    & $0.54(2)$    & $0.18(2)$     & $0.5$     & $-0.03(5)$   &  $0.07/4$
\end{tabular}
\end{ruledtabular}
\label{table:FK_xplus}
\end{table}

In this section, we focus on the numerical fits for the tricritical $Q$-state Potts model, i.e., the O$(n)$ loop model with
$n = \sqrt{Q}$ on the $x_+$ branch. Figures~\ref{fig7}(a)--\ref{fig7}(c) show the numerical data for $R_\text{FK}(\sqrt{L/2})$ and $R_\text{BB}(\sqrt{L/2})$ with $Q=1,2,3$, and Fig.~\ref{fig7}(d) corresponds to the critical four-state Potts model and will be discussed later. Similar to the $x_-$ branch, the data for $R_\text{FK}$ converge noticeably faster than those for $R_\text{BB}$, suggesting that the correction-to-scaling terms are weaker for $R_\text{FK}$ and therefore yield more precise fits.

We first consider the fits to the ratio $R_\text{FK}(\sqrt{L/2})$, summarized in Table~\ref{table:FK_xplus}. For the three cases $Q=1,2,3$, the fit is stable with varying correction term $y'$, and satisfactory fits are obtained with $L_\text{min}=32$. Thus, the fits reproduce the numerical data well, as shown in 
Figs.~\ref{fig7}(a)--\ref{fig7}(c).

We now turn to the fits of the ratio $R_\text{BB}(\sqrt{L/2})$, presented in Table~\ref{table:BB_xplus}. Here, we explored several values of the parameter $y'$, though for conciseness only representative results are shown. For small $Q$, such as $Q=1,2$, we tested a series of $y'$, and found that $R_\text{BB}$ is comparatively stable, the fluctuations in the fit results is of order $10^{-4}$. While for $Q=3$, the dependence on $y'$ is somewhat stronger, and the fluctuations of the result is of order $10^{-3}$. Nevertheless, the overall trends are already apparent from the values listed in Table~\ref{table:BB_xplus}.

\begin{table}
\caption{Fits of the data $R_\text{BB}(\sqrt{L/2})$ to the ansatz \eqref{eq-rjl} for $Q=1, 2$, and $3$ on the $x_+$ branch. Entries without errors represent fixed values in the corresponding (biased) fit. The fits with $L_\text{min}$ in boldface correspond to our preferred ones.}
\begin{ruledtabular}
\begin{tabular}{lrlllllr}
\multicolumn{1}{c}{$Q$}  &
\multicolumn{1}{c}{$L_{\text{min}}$}  &
\multicolumn{1}{c}{$R_\text{BB}$}  &
\multicolumn{1}{c}{$y$}  &
\multicolumn{1}{c}{$a_0$}  &
\multicolumn{1}{c}{$y'$}  &
\multicolumn{1}{c}{$a_1$}  &
\multicolumn{1}{c}{$\chi^2/\text{DF}$} \\
\hline
\multirow{6}{*}{$1$}
     &  $32$    & $1.3776(3)$    & $0.38(1)$    & $0.128(8)$    & $0.25$  & $-0.10(1)$   &  $0.09/5$  \\
     &  $64$    & $1.3776(6)$    & $0.38(3)$    & $0.13(2)$     & $0.25$  & $-0.10(4)$   &  $0.09/4$  \\
     &  $\mathbf{32}$    & $1.3770(4)$    & $0.33(1)$    & $0.087(3)$    & $0.5$   & $-0.067(9)$  &  $0.09/5$  \\
     &  $64$    & $1.3773(7)$    & $0.34(3)$    & $0.09(1)$     & $0.5$   & $-0.08(3)$   &  $0.08/4$  \\
     &  $32$    & $1.3762(4)$    & $0.290(9)$   & $0.072(2)$    & $1$     & $-0.11(2)$   &  $0.11/5$  \\
     &  $64$    & $1.3768(6)$    & $0.31(2)$    & $0.076(4)$    & $1$     & $-0.18(7)$   &  $0.08/4$  \\
\hline
\multirow{4}{*}{$2$}
     &  $32$    & $1.3403(4)$    & $0.35(1)$    & $0.16(1)$     & $0.25$   & $-0.06(2)$   &  $0.14/5$  \\
     &  $64$    & $1.3395(8)$    & $0.32(3)$    & $0.13(2)$     & $0.25$   & $-0.02(3)$   &  $0.09/4$  \\
     &  $32$    & $1.3399(4)$    & $0.326(9)$   & $0.128(4)$    & $0.5$    & $-0.04(1)$   &  $0.12/5$  \\
     &  $\mathbf{64}$    & $1.3393(7)$    & $0.31(2)$    & $0.119(9)$    & $0.5$    & $-0.01(2)$   &  $0.09/4$  \\
\hline
\multirow{6}{*}{$3$}
     &  $32$    & $1.2968(6)$    & $0.28(1)$    & $0.126(8)$    & $0.35$  & $0.03(1)$    &  $0.17/5$  \\
     &  $64$    & $1.2949(9)$    & $0.24(2)$    & $0.101(9)$    & $0.35$  & $0.07(1)$    &  $0.05/4$  \\
     &  $\mathbf{128}$   & $1.293(3)$     & $0.20(5)$    & $0.08(2)$     & $0.35$  & $0.10(3)$    &  $0.03/3$  \\
     &  $32$    & $1.2971(5)$    & $0.294(9)$   & $0.135(5)$    & $0.5$   & $0.02(1)$    &  $0.19/5$  \\
     &  $64$    & $1.2957(6)$    & $0.27(1)$    & $0.120(5)$    & $0.5$   & $0.06(1)$    &  $0.05/4$  \\
     &  $128$   & $1.294(1)$     & $0.24(2)$    & $0.11(1)$     & $0.5$   & $0.10(3)$    &  $0.04/3$
\end{tabular}
\end{ruledtabular}
\label{table:BB_xplus}
\end{table}

\subsection{Critical four-state Potts model}   \label{sec:potts4}

\begin{table}
\caption{Fits of the data $R_\text{FK}(\sqrt{L/2})$ to the ansatz \eqref{eq-rjl} for $Q=4$. Entries without errors represent fixed values in the corresponding (biased) fit. The fits with $L_\text{min}$ in boldface correspond to our preferred ones.}
\begin{ruledtabular}
\begin{tabular}{rlllllr}
\multicolumn{1}{c}{$L_{\text{min}}$}  &
\multicolumn{1}{c}{$R_{\rm FK}$}  &
\multicolumn{1}{c}{$y$}  &
\multicolumn{1}{c}{$a_0$}  &
\multicolumn{1}{c}{$y'$}  &
\multicolumn{1}{c}{$a_1$}  &
\multicolumn{1}{c}{$\chi^2/\text{DF}$} \\
\hline
$32$   & $1.1892(1)$    & $0.50(2)$   & $0.09(1)$     & $0.5$  & $0.06(2)$    &  $1.09/5$  \\
$64$   & $1.1890(3)$    & $0.47(6)$   & $0.08(3)$     & $0.5$  & $0.10(5)$    &  $0.96/4$  \\
$32$   & $1.1893(1)$    & $0.53(1)$   & $0.111(5)$    & $1$    & $0.12(4)$    &  $1.20/5$  \\
$\mathbf{64}$   & $1.1891(2)$    & $0.50(3)$   & $0.10(1)$     & $1$    & $0.3(2)$     &  $0.97/4$
\end{tabular}
\end{ruledtabular}
\label{table:FK_Q=4}
\end{table}

\begin{table}
\caption{Fits of the data $R_\text{BB}(\sqrt{L/2})$ to the ansatz \eqref{eq-rjl} for $Q=4$. Entries without errors represent fixed values in the corresponding (biased) fit. The fits with $L_\text{min}$ in boldface correspond to our preferred ones.}
\begin{ruledtabular}
\begin{tabular}{rlllllr}
\multicolumn{1}{c}{$L_{\text{min}}$}  &
\multicolumn{1}{c}{$R_{\rm BB}$}  &
\multicolumn{1}{c}{$y$}  &
\multicolumn{1}{c}{$a_0$}  &
\multicolumn{1}{c}{$y'$}  &
\multicolumn{1}{c}{$a_1$}  &
\multicolumn{1}{c}{$\chi^2/\text{DF}$} \\
\hline
$32$  & $1.2208(6)$ & $0.251(5)$ & $0.1169(7)$ & -  & 0 & $13.00/6$ \\
$64$  & $1.219(1)$  & $0.235(7)$ & $0.114(1)$  & -  & 0 & $3.95/5$ \\
$128$ & $1.217(2)$  & $0.22(1)$  & $0.110(2)$  & -  & 0 & $1.13/4$ \\
$256$ & $1.215(3)$  & $0.20(2)$  & $0.107(4)$  & -  & 0 & $0.71/3$ \\
$\mathbf{512}$ & $1.210(9)$ & $0.17(5)$ & $0.100(7)$ & -  & 0 & $0.08/2$ \\
$32$ & $1.216(2)$ & $0.21(1)$ & $0.107(2)$ & $1$ & $0.12(3)$ & $1.21/5$ \\
$\mathbf{64}$ & $1.214(4)$ & $0.19(3)$ & $0.103(5)$ & $1$ & $0.20(9)$ & $0.52/4$ \\
$128$ & $1.212(7)$ & $0.18(5)$ & $0.100(9)$ & $1$ & $0.3(3)$ & $0.45/3$ \\
$\mathbf{32}$ & $1.173(8)$ & $0.05$ & $0.09(1)$ & $0.38(5)$ & $0.093(2)$ & $0.32/5$ \\
$\mathbf{32}$ & $1.199(5)$ & $0.1$ & $0.078(9)$ & $0.39(7)$ & $0.082(4)$ & $0.34/5$ \\
$\mathbf{32}$ & $1.209(2)$ & $0.15$ & $0.088(6)$ & $0.5(1)$ & $0.07(1)$ & $0.45/5$
\end{tabular}
\end{ruledtabular}
\label{table:BB_Q=4}
\end{table}

\begin{table}
\caption{Fits of the data $R_\text{BB}(\sqrt{L/2})$ to the ansatz \eqref{def_ansatz_log} for $Q=4$. Entries without errors represent fixed values in the corresponding (biased) fit. The fits with $L_\text{min}$ in boldface correspond to our preferred ones.}
\begin{ruledtabular}
\begin{tabular}{rllllr}
\multicolumn{1}{c}{$L_{\text{min}}$}  &
\multicolumn{1}{c}{$R_\text{BB}$}  &
\multicolumn{1}{c}{$y$}  &
\multicolumn{1}{c}{$a_0$}  &
\multicolumn{1}{c}{$c$}  &
\multicolumn{1}{c}{$\chi^2/\text{DF}$} \\
\hline
$\mathbf{32}$  & $1.18(3)$   & $0.8(5)$   & $0.3(3)$    & $2(2)$ & $0.44/5$ \\
$32$           & $1.168(5)$  & $0.57(3)$  & $0.2394(8)$ & $1$    & $0.65/6$ \\
$\mathbf{64}$  & $1.170(7)$  & $0.59(5)$  & $0.240(2)$  & $1$    & $0.35/5$ \\
$\mathbf{32}$  & $1.184(3)$  & $0.82(4)$  & $0.347(9)$  & $2$    & $0.44/6$ \\
$\mathbf{32}$  & $1.193(3)$  & $1.08(4)$  & $0.58(3)$   & $3$    & $0.70/6$
\end{tabular}
\end{ruledtabular}
\label{table:BB_Q=4_log}
\end{table}

In this section, we discuss the fits for the critical O$(2)$ loop model, which corresponds to the critical four-state Potts model. At this point, the $x_-$ and $x_+$ branches of the O$(n)$ loop model merge. The numerical data are shown in 
Figs.~\ref{fig7}(d1) and \ref{fig7}(d2). It is evident that $R_\text{FK}$ still converges more rapidly toward its asymptotic limit than $R_\text{BB}$, implying that reliable fits can be expected for the former but not necessarily for the latter.

Table~\ref{table:FK_Q=4} summarizes the fits for $R_\text{FK}(\sqrt{L/2})$. The convergence is good, with a small error bar: $R_\text{FK}=1.1891(2)$. In this case, the statistical uncertainty is comparable to the variability of the estimates when $L_\text{min}$ and/or $y'$ are varied.

By contrast, $R_\text{BB}(\sqrt{L/2})$ converges much more slowly, as seen in 
Fig.~\ref{fig7}(d1). To address this, we modified our usual fitting strategy. We first applied the ansatz \eqref{eq-rjl} with $a_1=0$. The corresponding results are listed in the upper part of Table~\ref{table:BB_Q=4}. As $L_\text{min}$ increases, both $R_\text{BB}$ and $y$ decrease. Our preferred fit corresponds to $L_\text{min}=512$, yielding $R_\text{BB}=1.210(9)$, $y=0.17(5)$, $\chi^2=0.08$, and $\text{DF}=2$. These error bars should be interpreted cautiously, as $R_\text{BB}(\sqrt{L/2})$ may still be far from its asymptotic regime even for the largest sizes considered ($R_\text{BB}=1.2351(3)$ for $L=4096$ and $R_\text{BB}=1.2322(4)$ for $L=8182$).

We next fixed $y'=1$, finding a similar but slightly weaker decreasing trend for both $R_\text{BB}$ and $y$, with differences within the error bars. The preferred fit in this case corresponds to $L_\text{min}=64$, as shown in Table~\ref{table:BB_Q=4}. From these two fits, we estimate $R_\text{BB}\approx1.21(1)$ and $y=0.18(5)$.

Given the slow convergence of the data in Fig.~\ref{fig7} (d), it is natural to ask whether smaller values of the correction-to-scaling exponent $y$ could yield better fits. To explore this, we performed additional fits with fixed $y=0.05,0.10,0.15$ and free $y'$. The corresponding results are shown in the lower part of Table~\ref{table:BB_Q=4}. In all cases, the fits with $L_\text{min}=32$ were satisfactory. For larger $y$, the fits became unstable and are not reported. As seen in the table, increasing $y$ from $0.05$ to $0.15$ raises $R_\text{BB}$ from $1.17$ to $1.21$ [see Fig.~\ref{fig7} (d2)], while other parameters remain nearly constant [e.g., $y'=0.4(1)$, $a_0=0.08(1)$, $a_1=0.08(1)$]. We thus take $R_\text{BB}=1.19(2)$ as our best estimate, where the quoted uncertainty reflects the variability of these fits with respect to $y$ and includes the earlier results.

In Ref.~\cite{Fang_22}, a logarithmic ansatz to fit the backbone exponent $d_{\rm B}$ is proposed, which slightly improved their results. We repeated the analysis of $R_\text{BB}(\sqrt{L/2})$ using the form
\begin{equation}
R_{\text{BB}}(\sqrt{L/2}) = R_\text{BB} + a_0 \left( \ln L + c \right)^{y}\,,  \label{def_ansatz_log}
\end{equation}
and also tested more complex versions of this ansatz. However, adding the leading correction-to-scaling term $\left( \ln L + c \right)^{y} L^{y'}$ produced unstable fits with large error bars. Table~\ref{table:BB_Q=4_log} presents the results for Eq.~\eqref{def_ansatz_log}. An unbiased fit to the full form yields results consistent with those from the power-law ansatz \eqref{eq-rjl}, although the constant $c$
has a large uncertainty. Fixing $c=1,2,3$ and refitting gives $R_\text{BB}=1.18(1)$, again consistent with previous estimates.

In conclusion, our numerical results do not allow us to determine unambiguously whether logarithmic corrections to scaling exist at $Q=4$. Larger system sizes would be required to resolve this question, which is beyond our current computational capacity.

\section{Discussion}   \label{sec:res}

We performed extensive Monte Carlo simulations of the O($n$) loop model to measure the universal three-point structure constants $R_\text{FK}$ (FK cluster) and $R_\text{BB}$ (backbone) for both the critical and tricritical $Q$-state Potts models with $Q=1, 1.5, 2, 2.5, 3, 3.5$, and $4$. The results for $R_\text{FK}$ (see Table~\ref{table:R} and Fig.~\ref{fig1}) show excellent agreement with the exact theoretical predictions (see the Appendix for a brief introduction for the computational process used in Refs.~\cite{Delfino_11,Picco_13}), confirming the reliability of our simulations and finite-size scaling analysis over a wide range of $Q$ and Coulomb-gas coupling $g$.

In the critical regime, we find that $R_\text{BB}$ is systematically larger than $R_\text{FK}$, indicating that the backbone is geometrically denser and exhibits stronger three-point connectivity correlations than the FK clusters. A strikingly different behavior is observed in the tricritical regime: as the data shown in Table~\ref{table:R}, $R_\text{BB}$ and $R_\text{FK}$ become indistinguishable within numerical uncertainty. This leads to the conjecture
\begin{equation}
R_\text{BB}(g) \;=\; R_\text{FK}(g)\,.
\label{conj.R.tri}
\end{equation}
We have tested this conjecture by directly comparing our estimates of $R_\text{BB}$ with the exact $R_\text{FK}$ values for some typical values of $g$ from $1$ to $4/3$ (excluding $g=1$ due to its larger statistical uncertainty). The agreement is again excellent, with $\chi^2 = 4.41$ for $\text{DF}=6$. These observations suggest that Eq.~\eqref{conj.R.tri} likely holds throughout the entire tricritical branch $g \in [1,3/2]$.

This coincidence between the FK and backbone structure constants is not entirely unexpected. Previous work~\cite{Deng_04,Fang_22} reported that, along the tricritical line, the backbone fractal dimension $d_{\rm B}$ coincides with that of the FK cluster $d_f$, indicating that both structures share the same universality class at tricriticality. Our results extend this correspondence to three-point correlations, suggesting that the geometric and connectivity aspects of FK and backbone clusters become indistinguishable at the tricritical point. A full theoretical understanding of why such an equivalence arises, and how it connects to the underlying CFT, remains to be elucidated.

We conclude with some remarks about the perspectives of finding an analytical expression for $R_{\rm BB}(g)$ in the critical regime.
The Coulomb gas approach has been very powerful in the 1980s for obtaining the critical exponents of loop models \cite{DifrancescoSaleurZuber87}
(see also Ref.~\cite{JacobsenSaleur19}). However, the three-point structure constants of operators that modify the weights of loops \cite{Ikhlef_16}---and
of which $R_{\rm FK}(g)$ is a special case \cite{Delfino_11,Picco_13}---challenge this framework, since they do not respect the charge conservation
(up to discrete shifts by screening charges) that is built into the Coulomb gas. Such structure constants require the technology of imaginary Liouville
field theory \cite{Zamolodchikov2005} (or the conformal loop ensemble \cite{AngCaiSunWu24} in probability theory).
But also imaginary Liouville field theory is challenged
by other three-point structure constants of operators that insert open loop segments, since their expression \cite{JacobsenNivesvivatRibaultRoux25},
found from conformal bootstrap considerations, is more general than the DOZZ formula (see the Appendix). Therefore it is presently unclear by which means
a formula for $R_{\rm BB}(g)$ could possibly be derived, in particular since the backbone
exponent has features that are foreign to all the above-mentioned field-theoretical methods.
In particular, the operator inserting a backbone does not have fixed
Kac indices as $g$ is varied (see the Appendix).

\begin{acknowledgments}
This work has been supported by the National Natural Science Foundation of China (under Grant No. 12275263), the Innovation Program for Quantum Science and Technology (under grant No. 2021ZD0301900), the Natural Science Foundation of Fujian Province of China (under Grant No. 2023J02032), and
the Agence Nationale de la Recherche through the grant CONFICA (under grant No. ANR-21-CE40-0003).
\end{acknowledgments}

\appendix
*\section{Three-point structure constant} %\label{sec-app}

The three-point structure constant associated with the FK clusters in 2D, denoted as $R_{\rm FK}$, has been derived analytically within the framework of CFT in Refs.~\cite{Delfino_11,Picco_13}. In the following, we first review the basic CFT concepts and notations associated with the Potts model, and subsequently present how to calculate $R_{\rm FK}$ in the manner of Refs.~\cite{Delfino_11,Picco_13}.

\subsection{Critical Potts model} \label{sec-a-cp}

In 2D, the fixed point of the critical $Q$-state Potts model is described by a CFT with central charge
\begin{equation}
c(g) \;=\; 1 - 6\frac{(1-g)^2}{g}\,,  \label{def_c_Potts}
\end{equation}
where $g$ is the coupling constant, and relates to $Q$ via Eq.~\eqref{eq:Q_vs_g} within the range $g \in[1/2,1]$. For convenience, one may parametrize $g$ as
\begin{equation}
g \;=\; \frac{p}{p+1}, \label{def_g_vs_p}
\end{equation}
so that Eq.~\eqref{def_c_Potts} becomes
\begin{equation}
c(p) \;=\; 1 - \frac{6}{p(p+1)}\,. \label{def_c_Potts_p}
\end{equation}
The conformal weights of primary fields are then given by the Kac formula
\begin{equation}
\Delta_{m,n}(p) \;=\; \frac{ [(p+1)m -pn]^2 -1 }{4 p (p+1) }\,,    \label{def_Deltamn}
\end{equation}
which agrees with the convention of Ref.~\cite{Picco_13}, but differs by a factor of $2$ from Refs.~\cite{Delfino_11,Delfino_13}
which use the notation $\Delta_{m,n}(p)$ for the full scaling dimension (the sum of holomorphic and antiholomorphic conformal weights).

The scaling dimension associated with the FK cluster spin operator $\Phi_{\frac{1}{2},0}$ is
\begin{equation}
\Delta_\text{FK} \;=\; 2 \Delta_{\frac{1}{2},0}  \;=\;  
\frac{(p+3)(p-1)}{8p(p+1)} \,.  \label{def_DeltaFK}
\end{equation}
Rewriting Eq.~\eqref{def_DeltaFK} in terms of $g$ yields
\begin{equation}
\Delta_\text{FK}(g) \;=\; 1 - \frac{3}{8g} - \frac{g}{2}\,. \label{def_DeltaFK_g}
\end{equation}
This result is consistent with $y_{h1}$ given by Eq.~\eqref{eq-yh1} via the relation $\Delta_\text{FK}(g) =2- y_{h1}(g)$.
In the same way that $y_{h1}(g)$ is related to the Kac weight $\Delta_{\frac12,0}$, it is found that $y_{h2}(g)$ is related to
$\Delta_{\frac32,0}$, and $y_{t1}(g)$ to $\Delta_{2,1}$, and $y_{t2}(g)$ to $\Delta_{3,1}$. More generally, the full set of order-parameter
fields have weights $\Delta_{\frac12+\mathbb{N},0}$, and the full set of thermal fields have weights $\Delta_{2+\mathbb{N},1}$ \cite{DifrancescoSaleurZuber87,JacobsenSaleur19} (here $\mathbb{N}$ denotes the set of \emph{nonnegative} 
integers).

In CFT, each spin operator $\Phi_{m,n}$ is also associated with a charge $\alpha_{m,n}$, related to its scaling dimension $\Delta_{m,n}$ through
\begin{equation}
\Delta_{m,n} \;=\; \alpha_{m,n}(\alpha_{m,n} - 2\alpha_0)\,,  \label{def_Delta_alpha}
\end{equation}
where $2\alpha_0$ denotes the background charge,
\begin{equation}
2\alpha_0(g) \;=\; \sqrt{g} - \frac{1}{\sqrt{g}}\,. \label{def_2alpha0}
\end{equation}
Accordingly, the charge $\alpha_\text{FK}$ associated with the operator $\Phi_{\frac{1}{2},0}$ is obtained as the positive root of
$\alpha(\alpha-2\alpha_0)= \Delta_{\frac{1}{2},0}$, namely 
\begin{equation}
\alpha_\text{FK} \;=\; \frac{\sqrt{g}}{2}- \frac{1}{4\sqrt{g}} \,.
\label{def_alpha_FK}
\end{equation}

\subsection{Tricritical Potts model} \label{sec:Potts:tri}

The tricritical $Q$-state Potts model also admits a CG with coupling constant $g$ related to $Q$ by Eq.~\eqref{eq:Q_vs_g}, but within the range $g \in[1,3/2]$. In this sense, the tricritical model can be viewed as the analytic continuation of the critical Potts model, and vice versa.

Consequently, all eigenvalue expressions for the tricritical Potts model can be obtained as analytic continuations in $g$ of those for the critical case. Specifically, the associated charge $\alpha_\text{FK}$ can likewise be determined from Eq.~\eqref{def_alpha_FK}. However, the identification of operators differs from that of the critical Potts model.

First, the parametrization of the coupling strength $g$ in terms of $p$ is no longer given by Eq.~\eqref{def_g_vs_p}; instead, it becomes
\begin{equation}
g \;=\; \frac{p+1}{p} \,. \label{def_g_vs_p_tricrit}
\end{equation}
It is worth noting that, under this parametrization, the central charge $c(p)$ remains expressed by Eq.~\eqref{def_c_Potts_p}.

In the tricritical Potts model, the role of the spin operator corresponding to FK cluster is now played by $\Phi_{0,\frac{1}{2}}$ (rather than $\Phi_{\frac{1}{2},0}$ in the critical case), leading to
\begin{equation}
\Delta_\text{FK} \;=\; 2 \Delta_{0,\frac{1}{2}} \;=\; \frac{(p-2)(p+2)}{8p(1+p)} \,.
\end{equation}
By expressing this in terms of $g$ through Eq.~\eqref{def_g_vs_p_tricrit}, one recovers the same form as Eq.~\eqref{def_DeltaFK_g} for the critical case. This consistency confirms that the tricritical and critical Potts models are indeed related by analytic continuation in $g$.

\subsection{\texorpdfstring{Numerical evaluation of $R_{\rm FK}$}{Numerical evaluation of RFK}}

As derived analytically in Refs.~\cite{Delfino_11,Picco_13}, the three-point structure constant associated with FK clusters can be expressed as
\begin{equation}
R_{\rm FK}(g) \;=\; C_{\rm FK} \, C(g,\alpha_{\rm FK})\,,   \label{eq-rfk}
\end{equation}
where $C_{\rm FK} = \sqrt{2}$ is a $g$-independent prefactor, and $\alpha_{\rm FK}$ is the charge corresponding to the FK spin operator introduced above.

The function $C(g,\alpha)$ appearing in Eq.~\eqref{eq-rfk} is given by the so-called imaginary DOZZ formula
\begin{equation}
C(g,\alpha) \;=\; \frac{W(g,\alpha)}{W(g,0)}\,,
\end{equation}
where~\cite{Zamolodchikov2005}
\begin{multline}
W(g,\alpha) \;=\; \\[2mm] 
\frac{ [\Upsilon_\beta(\beta-\alpha)]^3 \,
       \Upsilon_\beta(2\beta - \beta^{-1} - 3\alpha)}
     { \big[\Upsilon_\beta(\beta-2\alpha)\,
       \Upsilon_\beta(2\beta - \beta^{-1} - 2\alpha) \big]^{3/2}}\,,
\end{multline}
with $\beta \equiv \sqrt{g}$. The normalization factor is given by
\begin{equation}
W(g,0) \;=\;
\frac{[\Upsilon_\beta(\beta)]^{3/2}}
     {[\Upsilon_\beta(2\beta-\beta^{-1})]^{1/2}}\,.
\end{equation}
Here $\Upsilon_\beta(x)$ is Zamolodchikov's special function~\cite{Zamolodchikov1996}, defined for $0 < \Re(x) < \widetilde{Q}$ by
\begin{multline}
\log \Upsilon_\beta(x)
\;=\; \\ \int_0^\infty \frac{\mathrm{d}t}{t}
\left[
\frac{\big(\frac{\widetilde{Q}}{2} -x\big)^2}{e^t} -
\frac{\sinh^2 \!\left[\frac{t}{2}\!\left(\frac{\widetilde{Q}}{2}-x\right)\!\right]}
     {\sinh \frac{\beta t}{2}\, \sinh \frac{t}{2\beta}}
\right]\,, \label{eq-Upsilon}
\end{multline}
where $\widetilde{Q} = \beta + \beta^{-1}$, and outside the strip of convergence for the above integral by the shift equations
\begin{equation}
\frac{\Upsilon_\beta(x+\beta)}{\Upsilon_\beta(x)} \;=\; \gamma(\beta x)\, \beta^{1-2\beta x}\,.
\end{equation}
In this equation, the function $\gamma$ is defined as 
\begin{equation}
\gamma(x) \;=\; \frac{\Gamma(x)}{\Gamma(x+1)}\,,
\end{equation}
where $\Gamma(x)$ is the Euler gamma function. 
Notice also the relation
\begin{equation}
 \Upsilon_\beta(x) \;=\; \Gamma_\beta^{-1}(x) \Gamma_\beta^{-1}(\widetilde{Q}-x) \,,
\end{equation}
where $\Gamma_\beta(x)$ is the Barnes double gamma function \cite[Appendix~B.1]{Eberhardt2023}. 

In summary, for a given value of $g$, the numerical evaluation of $R_{\rm FK}(g)$ proceeds as follows:
\begin{enumerate}
    \item Determine the value of $\alpha_{\rm FK}$ via Eq.~\eqref{def_alpha_FK};
    \item Evaluate $W(g,\alpha_{\rm FK})$ using the definitions above and a numerical implementation of $\Upsilon_\beta(x)$;
    \item Compute the normalization factor $W(g,0)$;
    \item Finally, evaluate $R_{\rm FK}(g) = \sqrt{2}W(g,\alpha_{\rm FK}) / W(g,0)$.
\end{enumerate}
This procedure allows for the numerical determination of $R_{\rm FK}(g)$ for any desired value of $g$ within the critical or tricritical regimes. The results for several selected values of $g$ are listed in Table~\ref{table:R}, while the continuous curve of $R_{\rm FK}(g)$ for successive $g\in[0.6,1.4]$ is shown as the line in Fig.~\ref{fig3}.

Moreover, the above procedure can also be directly applied to compute the three-point structure constant $R_{\text{spin}}$ associated with the spin clusters~\cite{Cai2025}. The calculation follows exactly the same formalism and functional relations as for the FK clusters, with only a few modifications. First, the dual spin operator $\Phi_{0,\frac{1}{2}}$ should be used in Eqs.~\eqref{def_Deltamn} and~\eqref{def_Delta_alpha} to obtain the corresponding scaling dimension $\Delta_{\text{spin}}$ and CG charge $\alpha_{\text{spin}}$. Second, for $R_{\text{spin}}(g) = C_{\text{spin}}(g) C(g,\alpha_{\text{spin}})$, in analogy with Eq.~\eqref{eq-rfk} for FK clusters, the prefactor $C_{\text{spin}}(g)$ becomes $g$-dependent and takes the form~\cite{Cai2025}
\begin{equation}
C_{\text{spin}} \;=\; \sqrt{\frac{2g}{Q}} \, \frac{\sin(2\pi g)}{\sin(\pi / g)}\,.
\end{equation}

%\bibliography{ref}

\end{document}